\documentclass[a4paper]{article}

\pdfoutput=1
\usepackage[a4paper,hmargin=3.0cm]{geometry}
\usepackage{lmodern}
\usepackage{booktabs}
\usepackage[T1]{fontenc}
\usepackage{subfloat}
\usepackage[utf8]{inputenc}
\usepackage{babel}

\makeatletter

\usepackage{mathrsfs}   
\usepackage{slashed}     
\usepackage{url}
\usepackage{graphicx}
\usepackage{color} 
\usepackage{bm}
\usepackage[numbers,sort&compress]{natbib}
\usepackage{float}
\usepackage{multirow}
\usepackage{amsmath}
\usepackage{amssymb}
\usepackage{breqn}
\usepackage{subcaption}
\usepackage{caption}
\usepackage{braket}
\usepackage{mciteplus}
\usepackage{authblk}
\usepackage{diagbox}
\usepackage{stackengine}
\usepackage[skip=0pt]{caption} 
\usepackage[normalem]{ulem}

\usepackage{tikz}

\usetikzlibrary{decorations.markings}
\usetikzlibrary{decorations.pathmorphing}

\tikzset{
    vector/.style={decorate, decoration={snake}, draw},
    fermion/.style={draw=black, postaction={decorate},
        decoration={markings,mark=at position .55 with {\arrow[scale=1.,>=stealth]{>}}}},
    fermionbar/.style={draw=black, postaction={decorate},
        decoration={markings,mark=at position .58 with {\arrow[scale=1.,>=stealth]{<}}}},
    fermionline/.style={draw=black},
    gluon/.style={decorate, draw=black,
        decoration={coil,amplitude=4pt, segment length=5pt}},
    scalar/.style={dashed, draw=black, postaction={decorate},
        decoration={markings,mark=at position .55 with {\arrow[scale=1.,>=stealth]{>}}}},
    scalarbar/.style={dashed, draw=black, postaction={decorate},
        decoration={markings,mark=at position .58 with {\arrow[scale=1.,>=stealth]{<}}}},
    scalarline/.style={dashed,draw=black},
    fmassin/.style={draw=black, postaction={decorate},
    decoration={markings, mark=at position .75 with {\arrow[scale=1.,>=stealth]{>}}, mark=at position .30 with {\arrow[scale=1.,>=stealth]{<}}}},
}

\usepackage[colorlinks=true,linkcolor=redLinks,citecolor=greenLinks,urlcolor=redLinks, pdfborder={0 0 1}]{hyperref}
\definecolor{greenLinks}{rgb}{0, 0.6, 0} 
\definecolor{blueLinks}{rgb}{0, 0, 0.6}
\definecolor{redLinks}{rgb}{0.6, 0, 0}
\definecolor{blackLinks}{rgb}{0.77,0.12,0.23} 
\definecolor{eprintLinks}{rgb}{0.4, 0.4, 0.4}
\definecolor{journalLinks}{rgb}{0.6, 0, 0}

\newcommand{\Tr}{\operatorname{Tr}}
\newcommand{\p}{\perp}

\begin{document}

\title{
 Reassessing the Connection Between the Color Glass Condensate and TMD PDFs}
\author[]{Benjamin Guiot\thanks{benjamin.guiot@usm.cl} }
\affil[]{\it{\normalsize Departamento de F\'isica, Universidad T\'ecnica Federico Santa Mar\'ia; Casilla 110-V, Valparaiso, Chile}}

\renewcommand\Authands{ and }
\date{}

\maketitle

\begin{abstract}
Paying particular attention to the different types of gluons, collinear, soft and Glauber, we argue that, in its standard formulation, the CGC does not contain the collinear sector required for a meaningful comparison with TMD factorization. The CGC Wilson lines usually identified as part of the proton matrix element should instead be interpreted as the equivalent of a soft factor. Restoring the missing collinear exchanges, we obtain a new definition of small-$x$ PDFs from an explicit calculation of the DIS cross section using CGC Feynman rules. The resulting expression can be cast in a form resembling TMD factorization without being equivalent to it. Finally, we discuss the evolution of these small-$x$ PDFs with respect to the rapidity cutoff $\Lambda$ and the Bjorken variable $x_b$.
\end{abstract}
\newpage

\tableofcontents

\section{Introduction}
The all-orders proof of factorization for small-$x$ partons is certainly an undervalued challenge in perturbative QCD. While several formalisms exist, in particular the $k_t$-factorization \cite{Catani1990,Catani1991,Collins1991,Levin91}, Color-Glass Condensate (CGC) \cite{McLerran1994,McLerran1994a, Iancu2001}, or a more recent factorization obtained with the Soft-Collinear Effective Theory (SCET) \cite{Neill2023}, none of them reached the same level of general acceptance of collinear and TMD factorization theorems.
Additionally, several groups have also investigated the connections between these different formalisms, with, for instance, the ITMD framework \cite{Kotko2015,Hameren2023}, studies of evolution equations \cite{Caucal2025, Caucal2026}, or the relation between the CGC and TMD PDFs \cite{Dominguez2011}.\footnote{Here and in the following, we use the term TMD PDFs only for parton densities defined in the TMD factorization \cite{Ji2004, Ji2005, Collins2023}.}\\

Analyzing some of these formalisms with the language of standard factorization theorems is a complementary task that may help clarify the appropriate framework for small-$x$ physics. Avsar followed this line of investigation in Ref.~\cite{Avsar2012}. In the present work, we focus on small-$x$ PDFs, namely transverse-momentum-dependent parton densities constructed from the CGC. 

In particular, we revisit the relation between the CGC and TMD PDFs proposed in Ref.~\cite{Dominguez2011}, which has been followed by several subsequent works. In that reference, the authors show that a specific limit of the CGC leads to an object similar to the TMD gluon density, with the soft factor absent, at least at leading order. However, by paying particular attention to the gluon kinematics, we reach a different conclusion. The CGC correlators should rather form a soft factor, while the proton matrix element of the collinear quark or gluon operators must be provided by a collinear sector that is absent from standard CGC calculations, as argued in Sec.~\ref{SecColl}.\\

Our main results are a new definition of small-$x$ PDFs, Eq.~(\ref{eq:mytmd2}) and the corresponding expression for the cross section, Eq.~(\ref{eq:cgctmd}). The latter can be cast in a form sharing some similarities with the TMD factorization, but is not equivalent. Our claim is that the CGC as standardly formulated omits collinear exchanges that are present by construction in TMD factorization, and we show what a CGC calculation looks like once they are restored.\\

The remainder of the paper is organized as follows. In Secs.~\ref{sec:cgc} and \ref{Sec:tmd}, we review the relevant ingredients of the CGC effective theory and TMD factorization. In Sec.~\ref{secArg}, we revisit the standard construction of small-$x$ PDFs from the CGC. In Sec.~\ref{sec:newdef}, we identify the missing collinear sector and introduce our modified definition of the small-$x$ PDFs. The corresponding calculation for the wave function is presented in Sec.~\ref{sec:diagram}, for the specific case of a longitudinally polarized photon. In Sec.~\ref{sec:evolution}, we discuss the dependence of the resulting PDFs on the rapidity cutoff $\Lambda$ and on $x_b$. Finally, we discuss shortly several interesting points in Sec.~\ref{sec:discussion}, before concluding in Sec.~\ref{sec:conclusion}.

\section{SIDIS and the Color Glass Condensate \label{sec:cgc}}
The CGC is based on the idea that fast gluons, with $k^+>\Lambda P^+=\Lambda^+$, are the color sources of soft classical gluons with $k^+<\Lambda^+$. Here, $P^+$ is the large light-cone component of the proton four-momentum, and $\Lambda$ has the meaning of a rapidity cutoff. In this effective theory, fast gluons are integrated out, giving rise to the current \cite{Iancu2001}
\begin{equation}
    J^\mu_a = \delta^{\mu +}\rho_a(x^-,x_\perp), \label{eq:source}
\end{equation}
while classical gluons are solutions of the classical Yang-Mills equations sourced by $\rho$, see, e.g., \cite{Gelis2010,Iancu2002, Iancu2004} for an introduction on this effective theory. A central ingredient is the CGC Wilson line
\begin{equation}
    U_{\text{cgc}}(\bm{x}_\perp)= P\exp \left(ig\int_{-\infty}^{+\infty}dx^- A^{+,a}_{\text{cl}}(x^-,\bm{x}_\perp)[\rho]t^a \right), \label{wilson}
\end{equation}
where $P$ indicates a path ordering, $A_{\text{cl}}$ is the {\it classical} gluon field, and $t^a$ are the SU(3) generators in the fundamental representation. Finally, the CGC correlators are defined as 
\begin{equation}
    \langle \mathcal{O}\rangle_\Lambda=\int [\mathcal{D}\rho]W_\Lambda[\rho]\mathcal{O}[\rho], \label{CGCcorre}
\end{equation}
where the operator $\mathcal{O}$ is usually the trace of some Wilson lines, and $W_\Lambda[\rho]$ is the weight functional, performing the average over $\rho_a$.\\

\begin{figure}[h!]
\begin{center}
 \includegraphics[width=10pc]{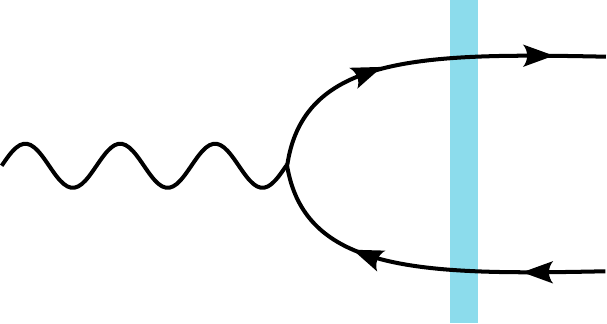}
\end{center}
\caption{DIS at leading order in the shockwave representation. The blue line represents multiple scattering, with at least one gluon exchanged with the target.\label{fig:cgc_dis}}
\end{figure}

For DIS, the leading-order (LO) diagram in the CGC framework is shown in Fig.~\ref{fig:cgc_dis}.
The virtual photon splits into a $q\bar{q}$ pair, and the vertical blue line corresponds to CGC Wilson lines implementing multiple gluon exchanges between the $q\bar{q}$ pair and the target. Multiple scattering resums the high gluon density, while the large $\log(x)$ are resummed by the Jalilian-Marian, Iancu, McLerran, Weigert, Leonidov, and Kovner (JIMWLK) evolution \cite{Balitsky1996,JalilianMarian1997,JalilianMarian1998,Kovchegov1999,Kovner2000,Iancu2001a,Iancu2001,Ferreiro2002,Weigert2002}  in $\Lambda$. The eikonal-CGC classical gluon is a particular type of Glauber gluon defined by the kinematics
\begin{equation}
    k^+k^- \ll k_\perp^2. \label{GG}
\end{equation}
Indeed, with $k^-=0$, the eikonal CGC conserving the minus component across the shockwave, the classical gluon is an extreme case of a Glauber gluon with $k^2=-k_\p^2$. 
As explained in the next section, multiple scatterings involving the exchange of soft and collinear gluons are partially handled by TMD factorization.

\section{TMD factorization and multiple scattering \label{Sec:tmd}}
\begin{figure}[h!]
\begin{center}
 \includegraphics[width=8pc]{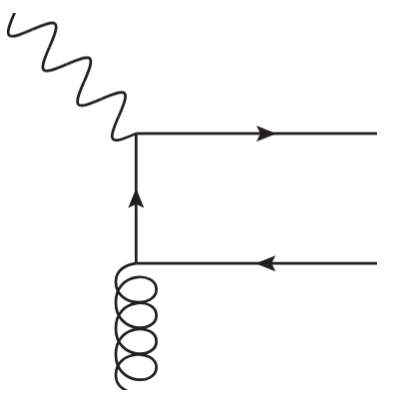}
\end{center}
\caption{Diagram corresponding to Fig.~\ref{fig:cgc_dis} in standard QCD. In both cases the target is not represented.\label{fig:nlo}}
\end{figure}

Fig.~\ref{fig:cgc_dis} should be compared to the next-to-leading-order (NLO) diagram presented in Fig.~\ref{fig:nlo}.
Even if not represented explicitly, Fig.~\ref{fig:nlo} includes multiple gluon exchanges between the outgoing partons and the target. The nature of these gluons is central, and a usual classification is
\begin{equation}
    \text{collinear:} \; p\sim Q(1, \lambda^2, \lambda), \quad  \text{soft:} \; p\sim Q(\lambda, \lambda, \lambda). \label{eq:scaling}
\end{equation}
Here, we are using the light-cone variable with the target having a large ``$+$'' component, and $\lambda=m/Q$, with $m$ a small mass and $Q$ the DIS variable.\\

Multiple exchanges of longitudinally polarized collinear and soft gluons with the target contribute at leading power and therefore have to be taken into account in TMD factorization. With the appropriate approximations, they can be factorized into two distinct Wilson lines. The collinear Wilson line\footnote{This is not a standard terminology, but in this manuscript, we will use the terms collinear, soft, and CGC Wilson lines.}, $U_c$, enters in the definition of collinear PDFs, schematically
\begin{equation}
    \text{PDF} = \langle P|\text{field } U_c \text{ field}|P\rangle.
\end{equation}
Here, $|P\rangle$ is the proton state. The TMD factorization also needs a soft factor, defined as a vacuum matrix element of two soft Wilson lines
\begin{equation}
    \text{soft factor} =  \langle 0|U_s^\dagger U_s|0\rangle.
\end{equation}
On the contrary, the glauber gluons cannot be factorized in this way, because their scaling, Eq.~(\ref{GG}), breaks the approximation required to turn the gluon propagator into an eikonal factor. However, in SIDIS, glauber gluons are not pinched, and a contour deformation brings them out of the glauber region. This last comment does not apply to the classical gluon discussed in Sec.~\ref{sec:cgc}, since, by definition, these gluons have $k^-=0$. What forces them to stay  in the Glauber region is not that their propagator is pinched between two poles, but rather the definition of the eikonal CGC.

\section{Standard construction of small-$x$ PDFs\label{secArg}}
We are ready to discuss the construction of small-$x$ PDFs given in \cite{Dominguez2011}. We present the main steps of the derivation, adopting in this section the notation of Ref.~\cite{Dominguez2011}. The starting point is dijet production in DIS, with $k_{1 \perp}$ and $k_{2 \perp}$ the transverse momentum of the outgoing jets.  The cross section obtained within the CGC effective theory reads

\begin{align}
\frac{d \sigma^{\gamma_{T, L}^* A \rightarrow q \bar{q} X}}{d^3 k_1 d^3 k_2}= & N_c \alpha_{e} e_q^2 \delta\left(p^+-k_1^+-k_2^+\right) \int \frac{d^2 x_1}{(2 \pi)^2} \frac{d^2 x_1^'}{(2 \pi)^2} \frac{d^2 x_2}{(2 \pi)^2} \frac{d^2 x_2^'}{(2 \pi)^2} \nonumber\\
& \times e^{-i k_{1 \perp} \cdot\left(x_1-x_1^'\right)} e^{-i k_{2 \perp} \cdot\left(x_2-x_2^'\right)} \sum_{\lambda \alpha \beta} \psi_{\alpha \beta}^{T, L \lambda}\left(x_1-x_2\right) \psi_{\alpha \beta}^{T, L \lambda *}\left(x_1^'-x_2^'\right) \nonumber\\
& \times\left[1+S_{x_g}^{(4)}\left(x_1, x_2 ; x_2^', x_1^'\right)-S_{x_g}^{(2)}\left(x_1, x_2\right)-S_{x_g}^{(2)}\left(x_2^', x_1^'\right)\right]. \label{xsecCGC}
\end{align} 
$S_{x_g}^{(2)}$ and $S_{x_g}^{(4)}$ are CGC correlators, defined in Eq.~(\ref{CGCcorre}), for 2 and 4 Wilson lines. Note also the change of notation $\Lambda \to x_g$ for the rapidity cutoff. In the correlation limit, where $q_\perp=|\bm{k}_{1 \perp}+\bm{k}_{2 \perp}|\ll P_t=|\bm{k}_{1 \perp}-\bm{k}_{2 \perp}|/2$, the expression above can be simplified after a change of variable and an expansion of the CGC Wilson lines to finally give
\begin{align}
\frac{d \sigma^{\gamma_T^* A \rightarrow q \bar{q} X}}{d \mathcal{P} . \mathcal{S} .}= & \alpha_{e} e_q^2 \alpha_s \delta\left(x_{\gamma^*}-1\right) z(1-z)\left(z^2+(1-z)^2\right) \frac{P_{\perp}^4+\epsilon_f^4}{\left(P_{\perp}^2+\epsilon_f^2\right)^4} \nonumber \\
& \times\left(16 \pi^3\right) \int \frac{d^3 v d^3 v^{\prime}}{(2 \pi)^6} e^{-i q_{\perp} \cdot\left(v-v^{\prime}\right)} 2\left\langle\operatorname{Tr}\left[F^{i-}(v) \mathcal{U}^{[+] \dagger} F^{i-}(v^{\prime}) \mathcal{U}^{[+]}\right]\right\rangle_{x_g},
\end{align}
with a similar equation for the longitudinal polarization. From the equation above, the small-$x$ (Weizs\"acker-Williams) PDF is extracted
\begin{equation}
    x_gG^{WW}(x_g,k_\perp)=\int d^3\xi_1 d^3\xi_2 e^{-ik_t . (\xi_{1\perp}-\xi_{2\perp})} \left\langle\operatorname{Tr}F^{i-}(\xi_1) \mathcal{U}^{[+] \dagger} F^{i-}(\xi_2) \mathcal{U}^{[+]}\right\rangle_{x_g}. \label{WW}
\end{equation}
This distribution is to be compared with TMD PDFs, written, with the help of translation invariance,
\begin{equation}
    xG(x, k_\perp)=\frac{1}{\braket{P|P} } \int d^3\xi_1 d^3\xi_2 \, e^{ixP^-(\xi_1^+-\xi_2^+) -ik_t . (\xi_{1\perp}-\xi_{2\perp})} \bra P \operatorname{Tr} F^{i-}(\xi_1) \mathcal{U}^{[+] \dagger} F^{i-}(\xi_2) \mathcal{U}^{[+]} \ket P \label{pdf1Simpl}
\end{equation}
The second to last step of the demonstration consists in taking the small-$x$ limit of Eq.~(\ref{pdf1Simpl}) by setting $x=0$ in the r.h.s, see also \cite{Marquet2016}, and accounting for the different normalizations used in both formalisms by the replacement
\begin{equation}
    \frac{\bra P ... \ket P}{\braket{P|P}}\longleftrightarrow \langle ... \rangle_{x_g}.\label{eqnorm}
\end{equation}
Finally, the last step is to set $x_g=x$, resulting is apparently two identical TMDs.\\

We do not share this conclusion, and explain in the next section why we believe that the CGC correlator should rather be interpreted as a CGC soft factor.

\section{A revisited definition of small-$x$ PDFs \label{sec:newdef}} 
\subsection{Missing collinear sector \label{SecColl}}

\begin{figure}[h!]
\begin{center}
 \includegraphics[width=18pc]{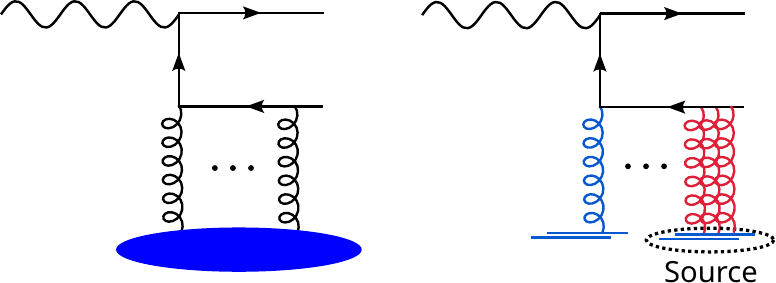}
\end{center}
\caption{On the left we show a general Feynman diagram for the interaction between the dipole and the target. For simplicity, only gluon exchanges with the antiquark are drawn. On the right, we show the same diagram, with collinear (in blue) and soft (in red) partons made explicit. \label{fig:cgc_coll}}
\end{figure}

In the CGC, the role of collinear partons is confined to the source term, Eq.~(\ref{eq:source}). But collinear gluons do not only radiate softer gluons. They also interact directly with the projectile, without the intermediary of soft gluons\footnote{Here and in the following we use the CGC convention where the terminology soft can also include Glauber gluons.}, as illustrated in Fig.~\ref{fig:cgc_coll}. This contribution has not been integrated out into the source term, and should consequently be considered explicitly when comparing the CGC with observables or quantities that receive contributions from collinear exchanges.

\begin{figure}[h!]
\begin{center}
 \includegraphics[width=12pc]{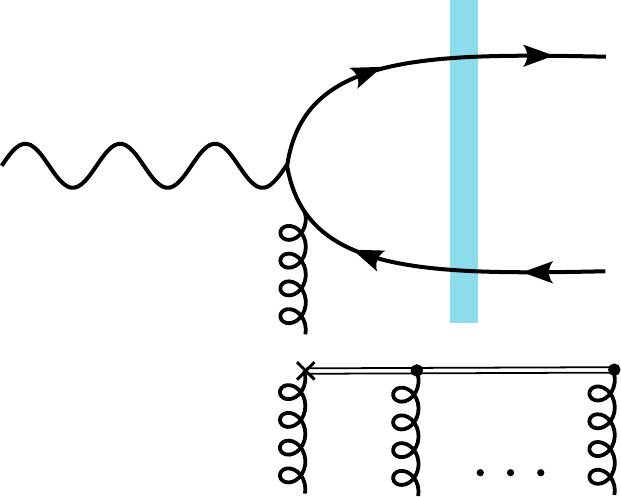}
\end{center}
\caption{DIS at leading order in the shockwave representation. The blue line represents multiple scattering, with at least one gluon exchanged with the target. The lower part of the diagram shows collinear gluons.\label{fig:cgc_mod}}
\end{figure}

Contrary to small-$x$ gluons, collinear gluons have a low occupancy number and can be treated with standard QCD. Then, we consider the following changes compared to Sec.~\ref{secArg}:
\begin{enumerate}
    \item We keep the CGC Wilson lines unexpanded. These Wilson lines give rise to the CGC soft factor (but it is fundamentally different from the standard TMD soft factor). 
    \item We keep one target-collinear gluon explicit, while the others build the collinear Wilson line. It is this Wilson line, and not the one made of soft gluons, that should enter into the proton matrix element of Eq.~(\ref{pdf1Simpl}).
    \item Diagrams with no collinear-gluon exchange cannot be related to the TMD factorization formula. Indeed, the latter is not valid when the initial parton is not collinear.
\end{enumerate}
In Fig.~\ref{fig:cgc_mod}, we show the diagram corresponding to points 1. and 2.
The collinear gluons can be factorized as usual and form the lower part of the diagram. The vertical gluon line ended by a cross is the gluon attaching to the hard region\footnote{The CGC is regularly applied to DIS data, including HERA data at moderate and large $Q^2$. Here we assume that the kinematics contain a hard scale $Q$, as required for a comparison with TMD factorization.}. Since, we treat collinear gluons with standard Feynman rules in momentum space, the bottom diagram gives the same contribution as in the TMD factorization.

\subsection{Small-$x$ PDFs, including the collinear sector \label{sec:construction}}
We want to compute the cross section associated to Fig.~\ref{fig:cgc_mod}, 
starting with the upper part of the diagram. If, for soft multiple scatterings, we keep using the CGC Feynman rules, and ignore the additional gluon, we obtain Eq.~(\ref{xsecCGC}). The collinear gluon introduces a new dependence on $k_t$ and $\xi/x_b$ in the wave function, whose detailed calculation is given in Sec.~\ref{sec:diagram}. Here, $\xi$ is the target large-momentum fraction carried by the exchanged collinear gluon. After the change of variable
\begin{equation}
    u=x_1-x_2, \quad v=zx_1+(1-z)x_2,
\end{equation}
with a similar operation for primed variables, the CGC cross section reads
\begin{multline}
\sigma^{\text{\scriptsize CGC}} \propto \int dz \frac{d^2 u}{(2 \pi)^2} \frac{d^2 u^{\prime}}{(2 \pi)^2} e^{i P_t \cdot (u^{\prime}-u)} \sum_{\lambda \alpha \beta} \psi_{\alpha \beta}^{\lambda^*}\left(z,u^{\prime},Q^2,\xi/x_b,k_t\right) \psi_{\alpha \beta}^\lambda\left(z,u,Q^2,\xi/x_b,k_t\right) \\
\times \int \frac{d^2 \mathbf{v}}{(2 \pi)^2} \frac{d^2 \mathbf{v}^{\prime}}{(2 \pi)^2} e^{i p_t \cdot\left(\mathbf{v}^{\prime}-\mathbf{v}\right)}\left[\frac{1}{2N_c}+\mathcal{S}_{\Lambda}^{(4)}(x_1, x_2 ; x_2^', x_1^')-\mathcal{S}_{\Lambda}^{(2)}\left(x_1, x_2\right)-\mathcal{S}_{\Lambda}^{(2)}(x_2^',x_1^')\right], \label{eq:cgcuv}
\end{multline}
with $p_t$ and $P_t$ the transverse momenta of the $q\bar{q}$ system and individual quark, respectively. To compare with the TMD factorization, one should integrate over $P_t$, giving a delta function which sets $u'=u$,
\begin{multline}
\sigma^{\text{\scriptsize CGC}} \propto \int d^2u  \left[\int dz \sum_{\lambda \alpha \beta} \left| \psi_{\alpha \beta}^\lambda\left(z,u,Q^2,\xi/x_b,k_t\right)\right|^2\right]
\int \frac{d^2 v}{(2 \pi)^2} \frac{d^2 v^{\prime}}{(2 \pi)^2} e^{i p_t \cdot\left(v^{\prime}-v\right)} \\
\times\left[\frac{1}{2N_c}+\mathcal{S}_{\Lambda}^{(4)}(x_1, x_2 ; x_2^', x_1^') -\mathcal{S}_{\Lambda}^{(2)}\left(x_1, x_2\right)-\mathcal{S}_{\Lambda}^{(2)}(x_2^',x_1^')\right]_{u=u'}. \label{eq:cgc1}
\end{multline}
Using the results of Sec.~\ref{sec:diagram}, the expression in the large square bracket can be replaced by
\begin{equation}
    H(u,Q^2,\xi/x_b,k_\p)=h(Q^2)c(u,\xi/x_b,k_\p,Q^2). \label{eq:hardF}
\end{equation}
It is not possible to factorize completely the $Q^2$ dependence. $h(Q^2)\propto Q^2$ carries the entire power dependence, and $c$ depends on $Q$ only through the dimensionless product $|u|Q$ within the Bessel function $K_0$.
With a new change of variables, $v'-v=b$, we can write the previous equation in the form
\begin{equation}
    \sigma^{\text{\scriptsize CGC}} \propto h(Q^2)\int \frac{d^2 b}{(2 \pi)^2} e^{i p_t \cdot b}\int \frac{d^2 u}{(2 \pi)^2}  c(u,\xi/x_b,k_\p,Q^2)\mathcal{S}_{\text{CGC}}(b,u;\Lambda), \label{eq:diag3}
\end{equation}
where we defined the CGC soft factor
\begin{equation}
        \mathcal{S}_{\text{CGC}}(b,u;\Lambda)=\int \frac{d^2 v}{(2 \pi)^2} \left[\frac{1}{2N_c}+\mathcal{S}_{\Lambda}^{(4)}(x_1, x_2 ; x_2^', x_1^') -\mathcal{S}_{\Lambda}^{(2)}\left(x_1, x_2\right)-\mathcal{S}_{\Lambda}^{(2)}(x_2^',x_1^')\right]_{u=u'}.
\end{equation}
The quadrupole and dipole are given by
\begin{equation}
    \mathcal{S}_{\Lambda}^{(4)}(x_1, x_2 ; x_2^', x_1^')=\frac{1}{(N_c^2-1)N_c}\Tr\left(V(x_1)t^aV^\dagger(x_2)V(x_2')t^aV^\dagger(x_1')\right),
\end{equation}
and 
\begin{equation}
    \mathcal{S}_{\Lambda}^{(2)}\left(x_1, x_2\right)=\frac{1}{(N_c^2-1)N_c}\Tr\left(V(x_1)t^aV^\dagger(x_2)t^a\right).
\end{equation}

While, in Sec.~\ref{secArg}, the dipole and quadrupole were expanded in order to build a correlator of two gluon fields, in the following, we keep them as they are, giving the CGC version of the soft factor.\\

The result for the reduced cross section, including the whole diagram of Fig.~\ref{fig:cgc_mod}, is an expression sharing similarities with the TMD factorization,
\begin{eqnarray}
    \sigma_{T,L}(x_b,Q^2,p_\p') &\propto & h_{T,L}(Q^2;\mu)\int d^2k_\p d^2p_\p d^2b \, e^{-ip_\p.b} f(x_b,k_\p,Q^2,b;\mu,\Lambda)\delta^2(k_\p+p_\p-p_\p') \nonumber \\
    &=& h_{T,L}(Q^2;\mu)\int d^2b \, e^{-ip_\p'.b} f(x_b,Q^2,b;\mu,\Lambda), \label{eq:cgctmd}
\end{eqnarray}
with $p_\p'$ the transverse momentum of the $q\bar{q}$ pair built from the initial $k_\p$ and soft rescatterings. In the second line of Eq.~(\ref{eq:cgctmd}), a simplification is obtained if $k_\p$ can be neglected in $c(u,\xi/x_b,k_\p,Q^2)$. For consistency, we restrict ourselves to the kinematic region where this holds, since this approximation is used in the first place to factorize the collinear gluon. While our notation mimics some of the functions defined in the TMD factorization, it is clear that they are not the same. For instance, the hard factor $h_{T,L}(Q^2;\mu)$ is not obtained by explicitly subtracting the soft and collinear regions from the general Feynman diagram. Another reason is that our small-$x$ PDFs, see Eq.~(\ref{eq:mytmd2}), are process dependent because the structure of Wilson lines depends on the process.\footnote{In the small-$x$ community, this statement is not necessarily surprising, see, for instance, Ref.~\cite{Kotko2015}.}\\

In Eq.~(\ref{eq:cgctmd}), we defined the small-$x$ PDFs
\begin{multline}
    f(x_b,k_\p,Q^2,b;\mu,\Lambda)=\int_0^1 \frac{d\xi}{\xi P^+} \int \frac{d^2 u}{(2 \pi)^2} c(u,\xi/x_b,k_\p,Q^2)\mathcal{S}_{\text{CGC}}(b,u;\Lambda) \\
    \times \int \frac{d\omega^-}{2\pi} \frac{d^2b'}{(2\pi)^2}e^{-i\xi P^+\omega^- -ik_\p\cdot b'}
    \bra P \operatorname{Tr}\Big[ F^{+i}\big(\tfrac{\omega}{2}\big) \,\mathcal{U}^{[+]}_{[\frac{\omega}{2},-\frac{\omega}{2}]}\, F^{+i}\big(-\tfrac{\omega}{2}\big)\, \mathcal{U}^{[+]}_{[-\frac{\omega}{2},\frac{\omega}{2}]}\Big]\ket P, \label{eq:mytmd}
\end{multline}
with the future-pointing staple
\begin{equation}
 \mathcal U^{[+]}_{[\frac{\omega}{2},-\frac{\omega}{2}]}
=\mathcal U_c^\dagger\left(\tfrac{\omega}{2};\infty\right)\,
\mathcal U^{T}\left[\tfrac{\omega_\perp}{2},-\tfrac{\omega_\perp}{2};z^-=+\infty\right]\,
\mathcal U_c\left(-\tfrac{\omega}{2};\infty\right),
\end{equation}
and $\omega=(0,\omega^-,b')$. Here $\mathcal U_c(x;\infty)\equiv\mathcal U^n[+\infty^-,x^-;x_\perp]$, where $\mathcal U^n$ is the Wilson line along the light-like direction $n^\mu$, with $n\cdot P=P^+$ and $n\cdot A=A^+$, and $\mathcal U^T$ is the transverse Wilson line at fixed $z^-$,
\begin{eqnarray}
    \mathcal U^n[a^-,c^-;x_\perp] = \mathcal P\exp\!\left[ig\int_{c^-}^{a^-}dz^-\,A^+(0^+,z^-,x_\perp)\right],\\
    \mathcal U^T[x_\perp,y_\perp;z^-] = \mathcal P\exp\!\left[-ig\int_{y_\perp}^{x_\perp}dz^i_\perp\,A^i(0^+,z^-,z_\perp)\right].
\end{eqnarray}

Neglecting $k_\p$ in $c(u,\xi/x_b,k_\p,Q^2)$, using the delta function and the complex exponentials, we obtain the second line of Eq.~(\ref{eq:cgctmd}) with
\begin{multline}
    f(x_b,Q^2,b;\mu,\Lambda)=\int_0^1 \frac{d\xi}{\xi P^+} \int \frac{d^2 u}{(2 \pi)^2} c(u,\xi/x_b,Q^2)\mathcal{S}_{\text{CGC}}(b,u;\Lambda) \\
    \times \int \frac{d\omega^-}{2\pi}e^{-i\xi P^+\omega^-}
\bra P \operatorname{Tr}F^{+i}\left(\frac{\omega}{2}\right)\,\mathcal U^{[+]}_{[\frac{\omega}{2},-\frac{\omega}{2}]}\,
F^{+i}\left(-\frac{\omega}{2}\right)\,\mathcal U^{[+]}_{[-\frac{\omega}{2},\frac{\omega}{2}]}\ket P, \label{eq:mytmd2}
\end{multline}
where $\omega=(0,\omega^-,b)$. The hypothesis leading to the final form of Eq.~(\ref{eq:cgctmd}) is correct as long  as we keep $k_\p\ll Q$, appropriate for the W-term of the TMD factorization.\\

We do not set $\Lambda=x_b$, and the collinear sector introduces a direct (nonperturbative) dependence on $x_b$. In principle, the scale $\mu$ should appear in $ f(x_b,Q^2,b;\mu,\Lambda)$ only after a proper treatment of the ultraviolet divergences. \\

In our opinion, this version is preferable to the one presented in Sec.~\ref{secArg} for several reasons:
\begin{enumerate}
    \item Soft gluons form the soft factor, absent in earlier derivations.
    \item Collinear-gluon exchanges are accounted for by collinear Wilson lines in the proton matrix element, similar to the TMD factorization.
    \item  Using the unclear relation (\ref{eqnorm}) is not necessary anymore.
\end{enumerate}

Finally, we note that Eq.~(\ref{eq:mytmd2}) shares some similarities with Eq.~(1.1) of Ref.~\cite{Neill2023}. It has a process-independent collinear function, the second line in Eq.~(\ref{eq:mytmd2}), and a process-dependent soft function ``analogous to the dipole function defined in the B-JIMWLK framework'' \cite{Neill2023}.

\section{Wave function with collinear insertion \label{sec:diagram}}
\subsection{Calculation}
Using the same conventions as Ref.~\cite{Caucal2021}, in particular the ``wrong'' light-cone gauge $A^-=0$, we compute the upper diagram of Fig.~\ref{fig:cgc_mod} for a longitudinal photon. We work in the reference frame where the photon transverse momentum is zero. The gluon shown explicitly is target collinear and does not interact with the shockwave. As long as the wave function is concerned, the gluon kinematics is approximated\footnote{This is the usual approximation used in collinear and TMD factorization for the hard coefficient.} by
\begin{equation}
    k\sim Q(1,\lambda^2,\lambda) \to \hat{k}= Q(1, 0, 0), \quad \lambda \sim \Lambda_{\text{QCD}}/Q, \label{eq:k_app}
\end{equation}
However, we will keep $k_\p$ explicit as long as possible, for a better control of approximations. The Feynman diagram represents a wave function with a collinear insertion followed by multiple scatterings of a dipole with the shockwave.\\

The amplitude reads
\begin{equation}
    \mathcal{M}=\frac{ee_f g}{2\pi}\int d^2x_\p d^2y_\p e^{-ip_{1\p}.x_\p}e^{-i(p_{2\p}-k_\p).y_\p}C_{\text{LO}}\mathcal{N}_{\text{LO}},
\end{equation}
with
\begin{equation}
    C_{\text{LO}}=V(x_\p)t^aV^\dagger (y_\p) - t^a, \label{eq:cr}
\end{equation}
and
\begin{equation}
    \mathcal{N}_{\text{LO}}=\bar{u}(p_1,\sigma)\int \frac{d^4 l}{(2\pi)^2} e^{il_\p.u} \gamma^- S^0(l)\slashed{\epsilon}(q,0)S^0(q-l)\slashed{\epsilon}(k,\lambda)S^0(k+q-l)\gamma^- v(p_2,\sigma')\delta(p_1^- - l^-).
\end{equation}
Similarly to Sec.~\ref{sec:construction}, we defined the variable $u=x_1-x_2$. Using $\slashed{\epsilon}(q,0)=Q/q^- \gamma^-$, and 
\begin{equation}
    \slashed{\epsilon}(k,\lambda)=\frac{\epsilon_\p^\lambda . k_\p}{k^-}\gamma^- -\epsilon_\p^\lambda . \gamma_\p \simeq \frac{\epsilon_\p^\lambda . k_\p}{k^-}\gamma^-, \label{eq:pol}
 \end{equation}
 where we used the fact that the ratio $k_\p/k^-$ is enhanced by the power of $\lambda^{-1}$, we obtain
 \begin{multline}
     \mathcal{N}_{\text{LO}}=\int \frac{d^4 l}{(2\pi)^2} e^{il_\p.u}\frac{N_{\text{LO}}\delta(p_1^- - l^-)}{[2l^+l^- -l_\p^2+i\epsilon][2(q^+-l^+)(q^- -l^-)-l_\p^2+i\epsilon]}\\
     \times \frac{1}{2(q^++k^+-l^+)(q^- +k^- -l^-)-(k_\p-l_\p)^2+i\epsilon}
 \end{multline}
 with
 \begin{equation}
     N_{\text{LO}}= 8Q \frac{l^-(l^- - q^-)^2}{q^-}\frac{\epsilon_\p^\lambda . k_\p}{k^-}\bar{u}(p_1,\sigma)\gamma^- v(p_2,\sigma').
 \end{equation}
 Note that we cannot neglect $q^+$ with respect to $k^+$ if we want to reach $Q^2$ of the order or larger than 1 GeV, which is common in fits on DIS data, and necessary for comparison with the TMD factorization. Using $\bar{u}(k,\sigma)\gamma^-v(p,\sigma')=2\sqrt{k^-p^-}\delta^{\sigma,-\sigma'}$ and $p_2^-=q^--l^-+k^-\simeq q^- -l^-=(1-z)q^-$, we write
 \begin{equation}
     N_{\text{LO}}=16Q\left(\sqrt{z(1-z)}q^- \right)^3(1-z)\frac{\epsilon_\p^\lambda . k_\p}{k^-}\delta^{\sigma,-\sigma'}
 \end{equation}

 We start with the contour integration on $l^+$. We close the path in the lower plane, picking the pole $l^+=l_\p^2/2l^-$, the two other poles lying in the upper half of the complex plane
  \begin{equation}
     \mathcal{N}_{\text{LO}}=\frac{-i\pi z N_{\text{LO}}}{zq^-}\int \frac{d^2 l_\p}{(2\pi)^2} e^{il_\p.u}\frac{1}{D_1 D_2},
 \end{equation}
 where we used $l^- = zq^-$. The expressions for the denominators are
 \begin{equation}
     D_1=-(1-z)Q^2\left(1-\frac{\xi}{x_b} \right) -\frac{1-z}{z}l_\p^2 - (k_\p-l_\p)^2,
\end{equation}
and
\begin{equation}
    D_2=(z-1)Q^2-\frac{l_\p^2}{z}
\end{equation}
In the equation above, we used
\begin{equation}
k^+=\xi P^+, \quad x_b=\frac{Q^2}{2q^-P^+}.    
\end{equation}
We use the Feynman parameterization to combine these denominators, leading to
\begin{equation}
    \mathcal{N}_{\text{LO}}=\frac{-i\pi z N_{\text{LO}}}{q^-}\int_0^1 dx \, e^{ixzk_\p.u}\int \frac{d^2 l_\p'}{(2\pi)^2} e^{il_\p'.u}\frac{1}{\left(l_\p^{\prime 2}+\Delta\right)^2},
\end{equation}
with 
\begin{equation}
   \Delta= z(1-z)Q^2\left(1- \frac{x\xi}{x_b}\right)+xz(1-xz)k_\p^2+i\epsilon
\end{equation}

Taking the Fourier transform, we finally obtain
\begin{equation}
     \mathcal{N}_{\text{LO}}=\frac{-i z N_{\text{LO}}}{4 q^-}|u|\int_0^1 dx \frac{e^{ixzk_\p.u}}{\sqrt{\Delta}}K_1(\sqrt{\Delta}|u|), \label{eq:nlo}
\end{equation}
with $K_1$ the modified Bessel function of the second kind.
We could have obtained a similar expression by using the rules for light-cone wave functions, with the exception that the incoming collinear gluon is not off-shell and has consequently no factor $1/k^-$.\\

In agreement with Eq.~(\ref{eq:k_app}), we now set $k_\p=0$ everywhere, except in the polarization vector for the reason explained after Eq.~(\ref{eq:pol}). Then, the equation above has the form
\begin{equation}
    \mathcal{N}_{\text{LO}}=\frac{-i z N_{\text{LO}}}{4 q^-}u^2\int_0^1 dx \frac{K_1(f(x))}{f(x)},
\end{equation}
with 
\begin{equation}
    f(x)=|u|\sqrt{z(1-z)Q^2\left(1- \frac{x\xi}{x_b}\right)+i\epsilon},
\end{equation}
leading to the result
\begin{equation}
    \mathcal{N}_{\text{LO}}=\frac{-i N_{\text{LO}}}{2(1-z)Q^2 q^-}\frac{x_b}{\xi}\left[K_0\left(|u|\sqrt{\Delta_1} \right) - K_0\left(|u|\sqrt{\Delta_0} \right)\right],
\end{equation}
with 
\begin{equation}
    \Delta_1 = z(1-z)Q^2\left(1-\frac{\xi}{x_b} \right), \quad \Delta_0 = z(1-z)Q^2.
\end{equation}
Note that $\Delta_1$ can be either positive or negative. In case of being negative, we can use $\sqrt{\Delta_1}=i\sqrt{|\Delta_1|}$ and
\begin{equation}
    K_0(i\sqrt{|\Delta_1|})=-\frac{\pi}{2}Y_0(\sqrt{|\Delta_1|})-\frac{i\pi}{2}J_0(\sqrt{|\Delta_1|}),
\end{equation}
where $J_0$ and $Y_0$ are the Bessel functions of the first and second kind.
Finally, squaring the wave function and summing on the quark antiquark helicities and averaging on the gluon polarization, we obtain
\begin{eqnarray}
    |\psi|^2&=&\frac{1}{2}\sum_{\sigma \sigma' \lambda}\mathcal{N}_{\text{LO}}^\dagger(z,u',Q^2) \mathcal{N}_{\text{LO}}(z,u,Q^2)\\
    &=&64z^3(1-z)^3(q^-)^2\frac{Q^2}{k_\p^2}\left[K_0\left(|u|\sqrt{\Delta_1} \right) - K_0\left(|u|\sqrt{\Delta_0} \right)\right] \nonumber\\
    &\times&\left[K_0^*\left(|u'|\sqrt{\Delta_1} \right) - K_0^*\left(|u'|\sqrt{\Delta_0} \right)\right]. \label{eq:psi2}
\end{eqnarray}
In this equation, we used the fact that $k_\p^2/(k^-)^2=Q^4\xi^2/(k_\p q^- x_b)^2$. The factor $1/k_\p^2$ is not part of the wave function, it is absorbed in the definitions (\ref{eq:mytmd}) and (\ref{eq:mytmd2}), using the relation between the field strength and the gluon field. Because of the arguments of the Bessel function, we cannot completely factorize the $Q^2$ dependence, and the function $h_{T,L}(Q^2)$ is simply proportional to $Q^2$. Finally, we note that Eq.~(\ref{eq:psi2}) has the typical CGC factor $z^3(1-z)^3$.\\

One could ask whether, applying the subtraction techniques of \cite{Collins2023} to define the hard factor, the $Q^2$ dependence would reside entirely in $h_{T,L}$. We leave this question open. It would not, however, make the resulting small-$x$ PDFs coincide with standard TMD PDFs: $\mathcal{S}_{\rm CGC}$ carries a target dependence, whereas the TMD soft factor is a vacuum matrix element and target independent. The two objects differ by construction.

\subsection{Analysis and limits}
We can rewrite
\begin{equation}
    \Delta_1=-z(1-z)(q+k)^2.
\end{equation}
It is proportional to the gluon-photon invariant mass, and $\Delta_1=0$ separates two different regions. When $\Delta_1$>0, $\xi<x_b$, and the system is spacelike. This region has no counterpart in collinear factorization, and we have the ordinary Bessel functions present in the CGC impact factors.  When $\Delta_1<0$, $\xi>x_b$, and the gluon-photon system is timelike. This region is compatible with the one in collinear factorization.\\

When $\xi \to 0$, $|\psi|^2\propto \xi^2$. Indeed, in this limit, $\sqrt{\Delta_1}=\sqrt{\Delta_0}[1-\xi/(2x_b)]$ and the square brackets of Eq.~(\ref{eq:psi2}) reduces to
\begin{equation}
 \left[K_0\left(|u|\sqrt{\Delta_1} \right) - K_0\left(|u|\sqrt{\Delta_0} \right)\right]  \simeq K_1(|u|\sqrt{\Delta_0})\frac{|u|\sqrt{\Delta_0}\xi}{2x_b},  
\end{equation}
where we used $K_0'=-K_1$. The two square brackets bring consequently a factor $\xi^2$. We can, consequently, safely integrate over $\xi$, from 0 to 1. The integrand is peaked at $\xi \sim x_b$ where $\Delta_1 \to 0$:
\begin{equation}
     \left[K_0\left(|u|\sqrt{\Delta_1} \right) - K_0\left(|u|\sqrt{\Delta_0} \right)\right]_{\xi \to x_b} \simeq K_0\left(|u|\sqrt{\Delta_1} \right) \simeq -\ln\left(|u|\sqrt{\Delta_1} \right).
\end{equation}
Here, we have neglected the terms that are not enhanced by a logarithm. When $\xi>x_b$, the contribution decreases for two reasons. In this region, $K_0(|u|\sqrt{\Delta_1})$ becomes a Hankel function whose modulus falls only like $(|u|\sqrt{|\Delta_1|})^{-1/2}$, and because of the proton matrix element.

\section{Evolution of small-$x$ PDFs\label{sec:evolution}}
\subsection{With $\Lambda$ \label{sec:evo_L}}
The CGC logic has not been modified, and the evolution of the correlators with $\Lambda$ is still driven by the JIMWLK equation. But the physical cross sections do not depend on renormalization scales. Then, the dependence of $f(x_b,Q^2,b;\mu,\Lambda)$ with $\Lambda$ depends on how this variable is split between $h_{T,L}(Q^2;\mu)$ and $c(u,\xi/x_b,k_\p,Q^2)$ beyond LO. These functions have been defined in Eqs.~(\ref{eq:hardF}), (\ref{eq:cgctmd}), and (\ref{eq:mytmd2}). A possible choice is to keep the form of Eq.~(\ref{eq:cgctmd}) close to the TMD factorization by keeping a simpler ``hard factor'' $h_{T,L}(Q^2,\mu^2)$, and by introducing all the dependence on $\Lambda$ in the function $c(u,\xi/x_b,k_\p,Q^2;\Lambda)$. In this case, the renormalization group equation is
\begin{equation}
    \frac{d}{d\Lambda}f(x_b,Q^2,b;\mu,\Lambda)=0, \label{eq:rgf}
\end{equation}
or equivalently
\begin{equation}
   \left(\frac{d}{d\Lambda}c(u,\xi/x_b,k_\p,Q^2;\Lambda)\right)\mathcal{S}_{\text{CGC}}(b,u;\Lambda)=- c(u,\xi/x_b,k_\p,Q^2;\Lambda)\left(\frac{d}{d\Lambda}\mathcal{S}_{\text{CGC}}(b,u;\Lambda)\right).
\end{equation}
But it is partly a matter of choice, and the dependence on $\Lambda$ could certainly be included in $h_{T,L}$.

\subsection{With $x_b$ \label{sec:evo_xb}}
It is often stated that small-$x$ PDFs obey a small-$x$ evolution equation in
$x_b$. On the contrary, we see from Eqs.~(\ref{eq:cgctmd}) and (\ref{eq:mytmd2}) that the dependence of the DIS cross section and PDFs on $x_b$ is not solely related to small-$x$ evolution. Indeed, the variable $x_b$ has two different origins: the collinear sector,
unrelated to small-$x$ evolution, and the lower limit of the projectile's
longitudinal phase space, fixed by its kinematics, which generates logarithms
$\ln(\Lambda/x_b)$ in the wave function at NLO and beyond. The choice
$\Lambda\sim x_b$ removes these large logarithms from the wave function and
allows their resummation by the weight functional and small-$x$ evolution
equations.\\

This is consistent with seminal works on $k_t$-factorization
\cite{Catani1990,Collins1991,Catani1994}. There, the authors start from
collinear factorization and resum large $\log(1/x)$ with the BFKL equation
\cite{Fadin1975,Kuraev1976,Kuraev1977,Balitsky1978}. In Mellin space, the
unintegrated gluon density obtained by these authors reads
\begin{equation}
    F_N(k_\p; \mu) = f_N(\mu)\,C_N(k_\p,\mu), \label{eq:updf}
\end{equation}
where $f_N(\mu)$ is the Mellin transform of the collinear gluon density, and
\begin{equation}
    C_N(k_\p,\mu)=\frac{\gamma_N(\alpha_s)}{k_\p^2}
    \left(\frac{k_\p^2}{\mu^2}\right)^{\gamma_N(\alpha_s)}, \label{eq:resum_fac}
\end{equation}
is the resummation factor, obtained from the BFKL equation.
Eqs.~(\ref{eq:updf}) and (\ref{eq:mytmd2}) have a similar structure: the
final distribution combines a collinear sector with a factor obeying a
small-$x$ evolution equation. Its dependence on $x$ is therefore not
determined by small-$x$ evolution alone, but also requires a non-perturbative,
$x$-dependent input from the collinear sector.

\section{Discussion \label{sec:discussion}}

\paragraph{DGLAP or BK?} 
In \cite{Albacete2012}, the authors conclude that their results beg ``for the development of a framework in which both $Q^2$ (DGLAP) evolution and $x$ (rcBK) non-linear evolution are jointly accounted for''. It is exactly what Eq.~(\ref{eq:mytmd2}) does. The collinear sector (second line) needs the usual renormalization procedure leading to the DGLAP equation, hence the renormalization scale $\mu$ in the l.h.s. In DIS, we usually set $\mu=Q$. At the same time, the small-$x$ PDF, or rather one of its components, obeys the JIMWLK equation, see Sec.~\ref{sec:evo_L}.

\paragraph{Consistency with collinear resummation of the BK kernel:} 
Finally, we believe that the literature on collinear resummation of BK kernel may support the present work. Without this resummation, higher-order corrections to small-$x$ kernels lack convergence, and can even lead to negative cross sections. It is interesting that Ref.~\cite{Iancu2015} says  explicitly that the collinear corrections arise from regions where non-linear effects are negligible. That sounds familiar to the collinear sector discussed in the core of this article. Indeed, if the dipole emits collinear gluons which interact with the target, these gluons can be factorized into the collinear Wilson line, see also Sec.~\ref{SecColl}.

\section{Conclusion \label{sec:conclusion}}
The main message of this work is that, even if a study focuses on small-$x$ formalisms applied to DIS or SIDIS, collinear exchanges between the virtual photon or final-state partons and the proton cannot always be neglected. This is particularly important for a meaningful comparison between small-$x$ PDFs, built from the CGC, and TMD PDFs. Indeed, the latter are obtained from the proof of TMD factorization where the initial parton is collinear. \\

We have computed the DIS cross section where soft gluons are described by the CGC Feynman rules, while collinear exchanges are treated in the standard way of factorization theorems. At the order of our calculation, the result can be written in a form mimicking some aspects of TMD factorization. However, the formulas obtained in these two formalisms are not equivalent, in particular, because what we called the CGC soft factor differs fundamentally from the one in TMD factorization.\\

From our cross-section we extract our definition of small-$x$ PDFs, Eq.~(\ref{eq:mytmd2}). They differ from the expressions given in the seminal work \cite{Dominguez2011} because the classical soft-gluon field is there identified with the operator entering the proton matrix element, whereas we argue that it plays a role closer to that of a soft factor. Our equations (\ref{eq:cgctmd}) and (\ref{eq:mytmd2}) are interesting in many respects: they propose, in our sense, a more natural definition of small-$x$ PDFs, as discussed at the end of Sec.~\ref{sec:construction}, possess both DGLAP and small-$x$ evolutions, and show similarities with the factorization formula of \cite{Neill2023}. One condition for their validity is the validity of the CGC formalism itself.\\

A further consequence concerns the $x_b$ dependence. In the eikonal CGC, this
dependence is entirely evolutionary. In our construction, $x_b$ also enters directly through the collinear sector, which introduces a non-perturbative dependence unrelated to
small-$x$ evolution. Consequently, the $x_b$ dependence of the small-$x$
PDFs of Eq.~(\ref{eq:mytmd2}) and of the DIS cross section is not determined
by small-$x$ evolution alone. As explained in Sec.~\ref{sec:evo_xb}, this is
consistent with the structure found long ago in $k_t$-factorization
\cite{Catani1990,Collins1991,Catani1994}.

\section*{Acknowledgments}
This work received support from Fondecyt (Chile) grants 1251322 and from ANID CCTVal CIA250027.

\bibliographystyle{BibFiles/t1}
\bibliography{BibFiles/GeneralBib}
\end{document}